\documentclass[twocolumn, prl, amssymb, superscriptaddress, aps,
preprintnumbers,amsmath,floatfix]{revtex4}
\usepackage{multirow}
\usepackage{isotope}
\usepackage{supertabular}
\usepackage{graphicx,amsfonts,color,epsfig}
\usepackage{amsmath}
\usepackage{amssymb}
\usepackage{url}

\begin{document}

\title{Theoretical predictions for the magnetic dipole moment of $^{229m}$Th}

\author{Nikolay \surname{Minkov}}
\email{nminkov@inrne.bas.bg} \affiliation{Institute of Nuclear Research and
Nuclear Energy, Bulgarian Academy of Sciences, Tzarigrad Road 72, BG-1784
Sofia, Bulgaria} \affiliation{Max-Planck-Institut f\"ur Kernphysik,
Saupfercheckweg 1, D-69117 Heidelberg, Germany}

\author{Adriana P\'alffy}
\email{Palffy@mpi-hd.mpg.de}
\affiliation{Max-Planck-Institut f\"ur Kernphysik, Saupfercheckweg 1,
D-69117 Heidelberg, Germany}


\date{\today}

\begin{abstract}
A recent laser spectroscopy experiment [J. Thielking {\it et al.}, Nature,
(London) {\bf 556}, 321 (2018)] has determined for the first time the
magnetic dipole moment of the 7.8 eV isomeric state $^{229m}$Th. The
measured value differs by a factor of approximately~5 from previous nuclear
theory predictions based on the Nilsson model, raising questions about our
understanding of the underlying nuclear structure. Here, we present a new
theoretical prediction based on a nuclear model with coupled collective
quadrupole-octupole and single-particle motions. Our calculations yield an
isomer magnetic dipole moment of $\mu_{\mbox{\scriptsize IS}}= -0.35\mu_N$
in surprisingly good agreement with the experimentally determined value of
$-0.37(6)\mu_N$, while overestimating the ground state dipole moment by a
factor 1.4. The model provides further information on the states' parity
mixing, the role and strength of the Coriolis mixing and the most probable
value of the gyromagnetic ratio $g_R$ and its consequences for the transition probability $B(M1)$.

\end{abstract}

\maketitle

{\it Introduction.}
The persistent interest of the metrology and atomic and nuclear physics
communities in the spectroscopic properties of the actinide nucleus
$^{229}$Th is related to the exceptionally low-energy $7.8$ eV isomeric state
$^{229m}$Th \cite{Beck_78eV_2007,Beck_78eV_2007_corrected}. Vacuum
ultraviolet (VUV) laser access to this nuclear state is believed to allow a
number of  applications such as a new ``nuclear clock'' frequency standard
\cite{Peik_Clock_2003,Peik_Clock_2015,Campbell_Clock_2012}, the development
of nuclear lasers in the optical range \cite{Tkalya_NuclLaser_2011} or a more
precise determination of the temporal variation of fundamental constants
\cite{Flambaum06,Berengut2009,Rellegert2010}. While so far direct laser
excitation of the isomer remains illusive due to the poor knowledge of the
exact transition frequency, recent experiments could demonstrate the
existence of the isomer \cite{Wense_Nature_2016}, and measure the isomer mean
half-life in neutral Th atoms \cite{Seiferle_PRL_2017}. In 2018, laser
spectroscopy experiments determined for the first time the  magnetic dipole
moment (MDM) of the nuclear isomeric state as $-0.37(6)\mu_N$
\cite{Thielking2018,Mueller18}, where $\mu_N$ stands for the nuclear
magneton. This breakthrough is particularly important because it opens the
possibility to experimentally probe  the nuclear isomeric state
via optical spectroscopy on the electronic hyperfine structure \cite{Beloy2014}. From the
nuclear structure point of view, however, the measured MDM value poses a
riddle: the only theoretical MDM prediction so far was providing the value
$\mu_{\mbox{\scriptsize IS}}= -0.076\mu_N$ \cite{Dyk98}, which is about five
times smaller than the measured value. This calls for new theoretical efforts
to understand the physical mechanism behind the Th isomer.

From the theoretical point of view, predictions on $^{229m}$Th have been
provided based on three approaches. Dykhne and Tkalya \cite{Dyk98} used the
Nilsson model \cite{Nilsson1955} to estimate the reduced transition
probability  $B(M1)$  and the isomer MDM, the former based on  Alaga rules
\cite{Alaga55}. Ruchowska and co-workers \cite{Ruch06} used the
quasiparticle-plus-phonon model to predict the reduced transition
probabilities  $B(M1)$ and $B(E2)$ for the isomeric transition. Finally, the
present authors have put forward a nuclear model approach that takes into
account the collective quadrupole-octupole vibration-rotation motion (typical
for the nuclei in the actinide region) the motion of the unpaired nucleon
within a reflection-asymmetric deformed potential with pairing correlations
and the Coriolis interaction between the single nucleon and the core
\cite{Minkov_Palffy_PRL_2017}.  This model  predicted the $B(M1)$ value for
magnetic decay of the isomer in the limits of $0.006-0.008$ Weisskopf units
(W.u.), well below earlier deduced values of 0.048 W.u. \cite{Dyk98,Tkalya15}
and 0.014 W.u. \cite{Ruch06}, thus potentially offering an explanation for
the recently reported experimental difficulties to observe the radiative
decay of the isomer \cite{Jeet_PRL_2015,Yamaguchi2015,LarsThesis2016}.

It is the purpose of this Letter to provide new nuclear structure predictions for the MDM of the $^{229}$Th ground and isomeric states by extending the  model approach in
Ref.~\cite{Minkov_Palffy_PRL_2017}. The  model parameters are independent of MDM
experimental data but rather rely on experimental energy levels and transition rates providing
a further test of the model's predictive capability. Based on different models for the collective
gyromagnetic ratio $g_R$, we obtain the isomeric-state MDM in the range
$\mu_{\mbox{\scriptsize IS}}=-0.25\mu_N$ to $-0.35\mu_N$, well in agreement with the
experimental values. The ground-state (g.s.) MDM value on the other hand is obtained in the
range $\mu_{\mbox{\scriptsize GS}}=(0.53-0.66)\mu_N$. This overestimates the latest
reported value of $0.36\mu_N$ deduced from state-of-the-art atomic structure calculations for
Th$^{3+}$ ions in Ref.~\cite{Safronova13} on the basis of electronic hyperfine splitting
data \cite{Campbell2011} or the older experimental result $\mu_{\mbox{\scriptsize
GS}}=0.45\mu_N$ \cite{Gerstenkorn74}. Our values for the g.s.
MDM are similar to the prediction $\mu_{\mbox{\scriptsize GS}}=0.54\mu_N$ based on a
modified Wood-Saxon potential \cite{Chasman1977}. Our calculations further show that as a
peculiarity, the single particle (s.p.) orbital for $^{229m}$Th features very strong parity
mixing. Furthermore, we find that whereas the effect of the Coriolis mixing is of crucial
importance for the existence and strength of the isomeric transition, it is relatively weak in the
g.s. MDM and negligible in the isomeric MDM. Finally, a comparison of our theoretical
prediction with the experimental results supports the consideration of a strong quenching of
the collective gyromagnetic ratio $g_R$ in $^{229}$Th and in consequence a lower reduced transition probability $B(M1)$ than so far assumed.

{\it Model approach.}
The model Hamiltonian is taken in the form \cite{Minkov_Palffy_PRL_2017}
\begin{eqnarray}
H=H_{\mbox{\scriptsize s.p.}}+H_{\mbox{\scriptsize pair}}+
H_{\mbox{\scriptsize qo}}+H_{\mbox{\scriptsize Coriol}}\, .
\label{Htotal}
\end{eqnarray}
Here, $H_{\mbox{\scriptsize s.p.}}$ is the s.p. Hamiltonian of the deformed
shell model (DSM) with a Woods-Saxon (WS) potential for axial quadrupole,
octupole and higher-multipo\-la\-rity deformations \cite{qocsmod} providing
the s.p. energies $E^{K}_{\mbox{\scriptsize sp}}$ with given value of the
projection $K$ of the total and s.p. angular momentum operators $\hat{I}$ and
$\hat{j}$, respectively, on the intrinsic symmetry axis.
$H_{\mbox{\scriptsize pair}}$ is the Bardeen-Cooper-Schrieffer (BCS) pairing
Hamiltonian \cite{Ring1980}. This DSM-BCS part provides the quasi-particle
(q.p.) spectrum $\epsilon^{K}_{\mbox{\scriptsize qp}}$ as implemented in
Ref.~\cite{WM10}. $H_{\mbox{\scriptsize qo}}$ represents a coherent QO motion
(CQOM) of the even--even core as considered in Refs.~\cite{b2b3mod,b2b3odd}.
$H_{\mbox{\scriptsize Coriol}}$ involves the Coriolis interaction between the
core and the unpaired nucleon (see Eq.~(3) in \cite{b2b3odd}). It is treated
as a perturbation with respect to the remaining part of (\ref{Htotal}) and
then incorporated into the QO potential of $H_{\mbox{\scriptsize qo}}$
\cite{NM13,Minkov_Palffy_PRL_2017}. The spectrum of (\ref{Htotal}) is then
obtained through the solution of the CQOM problem
\cite{b2b3mod,b2b3odd,MDSSL12} superposed to the q.p. spectrum obtained in
the DSM-BCS problem. It has the form of quasi-parity-doublets (QPD) ensuing
from the QO vibrations and rotations \cite{MDDSLS13} built on q.p. bandhead
(b.h.) states with given $K=K_{b}$ and parity $\pi^{b}$
\cite{NM13,Minkov_Palffy_PRL_2017}.

The Coriolis perturbed wave function
$\widetilde{\Psi}\equiv\widetilde{\Psi}^{\pi ,\pi^{b}}_{nkIMK_{b}}$
corresponding to the Hamiltonian (\ref{Htotal}) reads
\begin{equation}
\widetilde{\Psi}=\frac{1}{\widetilde{N}_{I\pi K_{b}}}
\left[\Psi^{\pi ,\pi^{b}}_{nkIMK_{b}} + A\sum_{\nu \neq b}
C^{I\pi}_{K_{\nu} K_{b}}\Psi^{\pi ,\pi^{b}}_{nkIMK_{\nu}}\right],
\label{wtcoriol}
\end{equation}
where $K_{\nu}= K_{b}\pm 1,\frac{1}{2}$; $C^{I\pi}_{K_{\nu} K_{b}}$ are
$K$-mixing coefficients, $\widetilde{N}_{I\pi K_{b}}$ is a normalization
factor and $A$ is the Coriolis mixing constant 
\cite{Minkov_Palffy_PRL_2017}. The unperturbed QO core-plus-particle wave
function in Eq.~(\ref{wtcoriol}) has the form \cite{NM13}
\begin{equation}
\begin{aligned}
&\Psi^{\pi ,\pi^{b}}_{nkIMK}(\eta ,\phi ,\theta)=
\frac{1}{N_{K}^{(\pi^{b})}}\sqrt{\frac{2I+1}{16\pi^2}}
\Phi^{\pi\cdot\pi^{b}}_{n k I} (\eta,\phi)\\
\times&\left[ D^{I}_{M\, K}(\theta )\mathcal{F}^{(\pi^{b})}_K+ \pi
\cdot\pi^{b}(-1)^{I+K}D^{I}_{M\,-K}(\theta)\mathcal{F}^{(\pi^{b})}_{-K}\right]\, ,
\end{aligned}
\label{wfpcore}
\end{equation}
where $D^{I}_{M\, K}(\theta )$ are the rotation functions of the three Euler
angles, $\Phi^{\pi\cdot\pi^{b}}_{n k I} (\eta,\phi)$ are the QO vibration
functions in radial ($\eta$) and angular ($\phi$) coordinates with
corresponding quantum numbers $n$ and $k$ (see \cite{MDSSL12,MDDSLS13} for
details) and $\mathcal{F}_{K_{b}}^{(\pi^{b})}$ is the parity-projected
component of the s.p. wave function of the b.h. state determined by DSM
\cite{qocsmod}. The quantity
$N_{K}^{(\pi^{b})}=\left[\left\langle\mathcal{F}_K^{(\pi^{b})}
\big|\mathcal{F}_K^{(\pi^{b})}\right\rangle \right]^{\frac{1}{2}}$ is the
corresponding parity-projected normalization factor. Note that the projection
is compulsory since for nonzero octupole deformation the s.p. wave function
obtained in DSM is parity mixed, i.e., it contains components possessing both
parities, $\mathcal{F}_{K}=\mathcal{F}_{K}^{(+)}+\mathcal{F}_{K}^{(-)}$. The
corresponding expectation value of the s.p. parity operator is then $-1\leq
\langle \hat{\pi}_{\mbox{\scriptsize sp}} \rangle =
\langle\mathcal{F}_{K}|\hat{\pi}_{\mbox{\scriptsize sp}}|
\mathcal{F}_{K}\rangle \leq 1$. The projection
$\mathcal{F}_{K_{b}}^{(\pi^{b})}$ is taken with respect to the experimentally
confirmed parity which is for the ground and isomeric states in $^{229}$Th
$\pi^{b}=+1$. In our case, the average $\langle \hat{\pi}_{\mbox{\scriptsize
sp}} \rangle$ is 0.4014 for the ground state and 0.0101 for the isomeric
state, respectively, showing that the parity mixing is very strong in the
DSM solution for $^{229m}$Th. We note that the parity-projection plus
renormalization procedure plays a considerable role in the resulting MDM
values.

The above CQOM-DSM-BCS formalism allows us to calculate the MDM in any state
of the spectrum by using the complete Coriolis perturbed wave function
(\ref{wtcoriol}). We consider the standard core plus particle magnetic dipole
($M1$) operator $\hat{M}1=\sqrt{3/4\pi}\mu_{N}\left [ g_{R}(\hat{I}-\hat{j})
+g_s\, \hat{s}+ g_l\, \hat{l} \right ]$, with $\hat{s}$ and $\hat{l}$ being
the operators of the s.p. spin and orbital momenta
($\hat{l}+\hat{s}=\hat{j}$), respectively, and $g_s$  the spin gyromagnetic
factor. The  MDM is determined by the matrix element $\mu =
\sqrt{\frac{4\pi}{3}}\langle \widetilde{\Psi}_{IIK}|\hat{M}1_{0}
|\widetilde{\Psi}_{IIK}\rangle$ where $\hat{M}1_{0}$ is the zeroth spherical
tensor component of $\hat{M}1$ taken after transformation to the intrinsic
frame (see Chapter 9 of Ref.~\cite{EG70}). Thus we obtain the following MDM
expression  for  a state with collective angular momentum $I$ and parity
$\pi$ built on a q.p. b.h. state with $K=K_{b}$ and $\pi =\pi^{b}$:

\begin{equation}
\begin{aligned}
\hspace{-0.45cm}
&\mu=\mu_{N}g_{R}I+\frac{1}{I+1}\frac{1}{\widetilde{N}_{I\pi K_{b}}^{2}}
\Bigg[K_{b}\frac{M^{\pi^{b}}_{K_{b}K_{b}}}{N^{(\pi^{b})}_{K_{b}}} \\
&+ 2A\, K_{b}\sum_{\substack {\nu \neq b\\   K_{\nu}= K_{b}=\frac{1}{2}}}
\delta_{K_{\nu}K_{b}}C^{I\pi}_{K_{\nu} K_{b}}
\frac{P^{b}_{K_{\nu}K_{b}} M^{\pi^{b}}_{K_{\nu}K_{b}}}
{N^{(\pi^{b})}_{K_{\nu}}N^{(\pi^{b})}_{K_{b}}} \\
& + A^{2}\mkern-18mu\!\!\sum_{\substack {\nu_{1,2} \neq b\\
 K_{\nu_{1},\nu_{2}}=K_{\nu} \\ =K_{b}\pm 1,\frac{1}{2}}}
\!\!\mkern-18mu\delta_{K_{\nu_{1}}\! K_{\nu_{2}}}\!K_{\nu}
C^{I\pi}_{K_{\nu_{1}}\! K_{b}}C^{I\pi}_{K_{\nu_{2}}\!K_{b}}
\frac{P^{b}_{K_{\nu_{1}}\!K_{\nu_{2}}}M^{\pi^{b}}_{K_{\nu_{2}}\!K_{\nu_{1}}}}
{N^{(\pi^{b})}_{K_{\nu_{1}}}N^{(\pi^{b})}_{K_{\nu_{2}}}}
\Bigg].
\end{aligned}
\label{mumod}
\end{equation}
%
The complete expression  would involve an additional so-called
decoupling term applying for the case $K_{b}=1/2$  \cite{EG70}, that we have disregarded
in Eq.~(\ref{mumod}), since it is not applicable for the states under
consideration here. Similarly, the second term in the brackets of
Eq.~(\ref{mumod}) is also not relevant for our problem. We use
$M^{\pi^{b}}_{K_{\mu}K_{\nu}}=
(g_l-g_R)K_{\nu}\delta_{K_{\mu}K_{\nu}}\langle\mathcal{F}_{K_{\mu}}^{(\pi^{b})}|
\mathcal{F}_{K_{\nu}}^{(\pi^{b})}\rangle
+(g_s-g_l)\langle\mathcal{F}_{K_{\mu}}^{(\pi^{b})}|\hat{s}_{0}|
\mathcal{F}_{K_{\nu}}^{(\pi^{b})}\rangle$, where $g_l=0\, (1)$ is the orbital
gyromagnetic factor for neutrons (protons), $g_s=0.6\ g_s^{\mbox{\scriptsize
free}}$ is the attenuated spin gyromagnetic factor with
$g_s^{\mbox{\scriptsize free}}=-3.826\, (5.586)$ for neutrons (protons)
\cite{Ring1980} and $g_R$ is the collective gyromagnetic factor which will be
discussed below. The factors $P^{b}_{K_{\nu_{1}}K_{\nu_{2}}}$ involve the BCS
occupation probabilities as shown in Ref.~\cite{Minkov_Palffy_PRL_2017}. The
third term in the brackets of Eq.~(\ref{mumod}) is important for the Coriolis
mixing. One can easily check that in the case of missing Coriolis mixing,
Eq.~(\ref{mumod}) appears in the usual form of the particle-rotor expression
\cite{Ring1980} and is consistent with the relevant limiting case.

{\it Numerical results.}
We calculate the $^{229}$Th ground and isomeric state  MDM  in Eq.~(\ref{mumod}) using the wave function (\ref{wtcoriol}). Following the procedure
described in Ref.~\cite{Minkov_Palffy_PRL_2017}, we have chosen the  quadrupole
($\beta_2$) and octupole ($\beta_3$) deformations entering DSM, the BCS pairing
parameters, the collective CQOM parameters and the Coriolis mixing strength such that  both
states $5/2^{+}$ g.s. and the isomeric $3/2^{+}$ form a quasi-degenerate pair and the
low-lying part of the $^{229}$Th spectrum \cite{ensdf} is well reproduced. This set of
parameters is the same as the one in Ref.~\cite{Minkov_Palffy_PRL_2017} except for the
Coriolis mixing constant $A$ which was slightly shifted from 0.158 to 0.184 keV. The latter is
due to fixing a minor numerical inaccuracy found in the Coriolis mixing coefficients. This leads to a negligible change in the other energies and slight change in the
transition rates discussed below.

The parameter that needs special consideration is the collective gyromagnetic factor $g_R$. In
Ref.~\cite{Minkov_Palffy_PRL_2017} the $B(M1)$ values were obtained by using the
phenomenological expression $g_R=Z/(Z+N)$ with $Z$ and $N$ the proton and neutron
numbers, respectively, adopted  on the basis of the liquid-drop-model \cite{Way39}.
However, it is well known that similarly to $g_s$ (which is typically attenuated by the
quenching factor 0.6 taking into account spin polarization effects \cite{Mottelson60}) in most
deformed nuclei  $g_R$  is also lowered by 20-30{\%} or more \cite{EG70}. This effect has
been proven both experimentally \cite{Boden62} and theoretically
\cite{NP61,PBN68,Greiner65}. We note that the experimental determination of $g_R$ is
model dependent and in fact usually involves knowledge of MDM. In theory the lowering of
$g_R$ is explained through a suppressed relative contribution of the proton system to the total
moment of inertia due to the pairing interaction \cite{NP61,PBN68}. Thus in odd-mass nuclei
the quenching of $g_R$ is stronger when the odd particle is a neutron and weaker when it is a
proton compared to the quenching in the adjacent even-even nuclei. In this basic approach
$g_R$ is usually calculated through the Inglis-Belyaev cranking procedure \cite{Iglis1,Iglis2}
using \cite{NP61,PBN68} the Nilsson deformed shell model \cite{Nilsson1955} or applying a
density-dependent Hartree-Fock plus BCS (HFBCS) model \cite{Sprung79}.

We introduce the quenching factor $q_{R}<1$, such that $g_R=q_{R}Z/(Z+N)$
with $Z/(Z+N)=0.393$ for $^{229}$Th. We have used the following  $g_R$
reference values: $g_R=0.28$ and $0.27$ for $^{228}$Th and $^{230}$Th,
respectively, obtained in the early cranked Nilsson approach \cite{NP61};
$0.24$ obtained for $^{230}$Th in the cranked HFBCS calculation
\cite{Sprung79}; and $0.31\pm 0.03$ deduced in Ref.~\cite{Ton70} on the basis
of experimental measurements for the ground state MDM and $M1/E2$ mixing
ratios for two ground state intraband transitions in $^{229}$Th. Keeping in
mind that for the odd-neutron nucleus $g_R$ is lower compared to the
neighboring even-even cases we consider the quenching factors $q_R=0.7$ and
0.6 based on the cranked Nilsson and HFBCS calculations, respectively, and
$q_R=0.8$ given by the experimental estimate \cite{Ton70}.

\begin{table*}[htbp]
\caption{Calculated ground state and isomer MDM
$\mu_{\mbox{\scriptsize GS}}$ and $\mu_{\mbox{\scriptsize IS}}$, respectively,
obtained for $^{229}$Th for several $q_R$ ($g_R$ quenching) values in comparison
with previous nuclear theory predictions and experimentally deduced MDM values.}
\begin{center}
{\small
\tabcolsep=5pt
\begin{tabular}{crrrrcccccccc}
\hline\hline
\multirow{2}{4em}{\ \ \ \ $\mu \, (\mu_N)$} &
\multicolumn{4}{c}{$q_R$ (this work)}& & \multicolumn{2}{c}{nuclear theory}
& & \multicolumn{4}{c}{laser spectroscopy} \\
\cline{2-5}  \cline {7-8} \cline{10-13}
& $1.0$ & $0.8$ & $0.7$ & $0.6$ & & Ref.~\cite{Chasman1977} & Ref.~\cite{Dyk98}
& & Ref.~\cite{Gerstenkorn74} &
Ref.~\cite{Safronova13}&
Ref.~\cite{Mueller18} &
Ref. \cite{Thielking2018}\\
\hline \\[-9pt]
$\mu_{\mbox{\scriptsize GS}}$ &   0.654 &  0.591 &  0.559 &  0.528 & &0.54&
--  & & 0.46(4) & 0.360(7) & -- & -- \\
$\mu_{\mbox{\scriptsize IS}}$ & $-0.253$&$-0.300$&$-0.323$&$-0.347$& &--  &
$-0.076$ & &-- & -- & $(-0.3)$--$(-0.4)$ & $-0.37(6)$ \\
\hline \hline
\end{tabular}
\label{tab:mdms} }
\end{center}
\end{table*}

The MDM results are shown in Table~\ref{tab:mdms} and compared with existing theoretical and
experimental values. The previous nuclear theory approaches  were based on the Nilsson
model \cite{Dyk98} and modified Woods-Saxon potential \cite{Chasman1977}. The
experimental data on the MDM stems from laser spectroscopy measurements of the electronic
hyperfine splitting of Th ions. The early work in Ref.~\cite{Gerstenkorn74} extracted the
hyperfine constants $A$ and $B$ from Th$^+$ spectra and deduced based on rather
simplified atomic structure matrix elements the ground state MDM of
$\mu_{\mbox{\scriptsize GS}}=0.45(4)\,\mu_N$. This value was corrected in Ref.
\cite{Safronova13} to $\mu_{\mbox{\scriptsize GS}}=0.360(7)\,\mu_N$ based on a more
recent measurement of the hyperfine structure of $^{229}$Th$^{3+}$ ions
\cite{Campbell2011} and state-of-the-art atomic structure calculations based on the
coupled-cluster model. The first experimental observation of the isomer electronic hyperfine
splitting in $^{229}$Th$^{2+}$ was reported only recently \cite{Thielking2018}. Based on
this measurement, an isomer MDM value of $\mu_{\mbox{\scriptsize IS}}=-0.37(6)\,\mu_N$
\cite{Thielking2018} or in the range of between $-0.30$ and $-0.40\,\mu_N$ \cite{Mueller18}
was extracted. Note that Ref.~\cite{Thielking2018} makes use of the ground state MDM value
$\mu_{\mbox{\scriptsize GS}}=0.360(7)\,\mu_N$ from Ref.~\cite{Safronova13} to extract
the experimental value for $\mu_{\mbox{\scriptsize IS}}$.

Regarding our model predictions, an overall observation is the decrease of
both MDMs $\mu_{\mbox{\scriptsize GS}}$ and $\mu_{\mbox{\scriptsize IS}}$
(the latter going towards more negative values) with the decrease of $g_R$.
As seen from Table~\ref{tab:mdms} we have obtained for the ground state MDM
the model values in the range between $\mu_{\mbox{\scriptsize GS}}=0.655$ and
$0.530\,\mu_N$ for $q_R$ taking the values 1.0, 0.8, 0.7 and 0.6. Comparing
them to the experimental values in Refs.~\cite{Gerstenkorn74} and
\cite{Safronova13} we see that they overestimate the first one,
$0.46(4)\,\mu_N$, by a factor between 1.42 and 1.15. The latter one is
overestimated by a factor between 1.82 and 1.47. On the other hand our values
for the isomer MDM which vary between $\mu_{\mbox{\scriptsize IS}}= -0.253$
and $-0.347\,\mu_N$  essentially corroborate the experimental values in
Refs.~\cite{Thielking2018} and \cite{Mueller18}. We see that the three values
obtained at $q_R=0.8$, 0.7 and 0.6 enter the error bar for the value of
$\mu_{\mbox{\scriptsize IS}}=-0.37(6)\,\mu_N$ in \cite{Thielking2018}. The
value at $q_R=1$  underestimates (in absolute value) by a factor 0.84  the lower limit
$-0.30\,\mu_N$ of the values in Ref.~\cite{Mueller18}.
Comparison with previous theoretical results shows that  our calculations
corroborate the g.s. MDM values $\mu_{\mbox{\scriptsize GS}}=0.54\,\mu_N$
obtained by Chasman et al \cite{Chasman1977} and disagree with the isomer MDM
value $\mu_{\mbox{\scriptsize IS}}= -0.076\,\mu_N$ obtained in \cite{Dyk98}. As a further check, the model predicts a spectroscopic electric quadrupole moment $Q_{GS}$=2.79 eb, close to the experimentally determined 3.11(16) eb \cite{Campbell2011} or 3.14(3) eb \cite{Bemis_pscr88}.

\begin{table}[htbp]
\caption{Predicted $B(M1)$ values (in W.u.) for several transitions involving
yrast (yr) and excited (ex) QPD states of $^{229}$Th obtained for four $q_R$
values in comparison with available experimental data. }
\begin{center}
{\small
\tabcolsep=4pt
\begin{tabular}{cccccc}
\hline\hline
\multirow{2}{4em}{Decay} &
\multicolumn{4}{c}{$q_R$}&
\multirow{2}{4em}{Exp.~\cite{nndc_gam}} \\ \cline{2-5}
&$1.0$ & $0.8$ & $0.7$ & $0.6$ &\\
\hline \\[-9pt]
$\frac{3}{2}^{+}_{\mbox{\scriptsize ex}}\, \rightarrow\,
\frac{5}{2}^{+}_{\mbox{\scriptsize yr}}$&
0.0081 & 0.0068 & 0.0062 & 0.0056& -- \\

$\frac{7}{2}^{+}_{\mbox{\scriptsize yr}}\, \rightarrow\,
\frac{5}{2}^{+}_{\mbox{\scriptsize yr}}$&
0.0096  &0.0043 & 0.0025 & 0.0011 & 0.0110 (40) \\

$\frac{9}{2}^{+}_{\mbox{\scriptsize yr}}\, \rightarrow\,
\frac{7}{2}^{+}_{\mbox{\scriptsize yr}}$&
0.0185  &0.0097 & 0.0065 & 0.0038 & 0.0076 (12) \\

$\frac{9}{2}^{+}_{\mbox{\scriptsize yr}}\, \rightarrow\,
\frac{7}{2}^{+}_{\mbox{\scriptsize ex}}$&
0.0144  & 0.0147& 0.0149 & 0.0151 & 0.0117 (14) \\
 \hline \hline
\end{tabular}
\label{tab:bm1trans} }
\end{center}
\end{table}

While the variation of $q_R$ ($g_R$) in Table I does not affect the energy levels and $B(E2)$
transition rates, it does change the predicted  $B(M1)$ transition rates  in
Ref.~\cite{Minkov_Palffy_PRL_2017}. This is illustrated in Table~\ref{tab:bm1trans}
where the model predictions for several $B(M1)$ values (including that of the isomer
transition) obtained for the different $q_R$ ($g_R$-quenching) values are given together with
the available experimental data. The transition rates in the first column (with $q_R=1.0$) are
slightly different from the original calculation in \cite{Minkov_Palffy_PRL_2017} due to the
already mentioned correction in the numerical mixing coefficients. We see that the
isomer-decay $B(M1)$ value (first row) gradually decreases with the decrease of $g_R$ in
consistence with the MDM behavior. Faster decrease of
$B(M1)$ with $g_R$ is observed for the yrast intraband transitions $7/2^{+}\rightarrow
5/2^{+}$ and $9/2^{+}\rightarrow 7/2^{+}$ whereas the B(M1) values for the yrast-excited
interband transition $9/2^{+}_{\mbox{\scriptsize yr}}\rightarrow 7/2^{+}_{\mbox{\scriptsize
ex}}$ practically remain unaffected. Comparison with the experimental data shows that the
first intraband transition $7/2^{+}\rightarrow 5/2^{+}$ which is originally underestimated by
the model goes even further down with $g_R$ by an order of magnitude. On the other hand
the other intraband transition $9/2^{+}\rightarrow 7/2^{+}$ approaches rather well the
corresponding experimental value for $q_R=0.7$.

The dependence in Tables \ref{tab:mdms} and \ref{tab:bm1trans} suggests that,
since  the quenching of $g_R$ is physically reasonable in the considered
limits, we may state that the model predictions for $\mu_{\mbox{\scriptsize
GS}}$ and $\mu_{\mbox{\scriptsize IS}}$ obtained with $g_R$ attenuated below
the $Z/(Z+N)=0.393$ value better approach the corresponding experimental
values with $\mu_{\mbox{\scriptsize IS}}$ firmly entering the uncertainty
bars. In addition, this result suggests that a further slight decrease of the
predicted isomer-decay $B(M1)$ value as compared to
Ref.~\cite{Minkov_Palffy_PRL_2017} towards $B(M1)=0.005$ W.u. is probable.

A word is due on the role of the Coriolis mixing for the calculated MDM values. The
attenuation of $g_R$ counteracts the mixing-effect on the connection between the two states.
For example our numerical analysis shows that a slight raising of the Coriolis constant $A$
from 0.184\,keV  used in this work to the value of 0.230\,keV leads to a slight decrease in
$\mu_{\mbox{\scriptsize GS}}$ at $q_R=1.0$ from 0.654 to $0.643\, \mu_N$ with a
negligible decrease in $\mu_{\mbox{\scriptsize IS}}$ in the fifth digit leaving the reported
value of $-0.253\, \mu_N$ unaffected. At the same time the isomer $B(M1)$ transition rate
raises from 0.0081 to 0.0121 W.u. The predictions for the $E2$ transitions are also affected
with the isomer $B(E2)$ raising from 29 to 43 W.u.  Conversely by setting the Coriolis mixing
strength zero as a limiting case, $\mu_{\mbox{\scriptsize GS}}$ slightly increases to
$0.677\,\mu_N$ whereas $\mu_{\mbox{\scriptsize IS}}$ remains again unaffected. This
analysis reveals some peculiarities of the present model mechanism in $^{229}$Th. Whereas
the Coriolis mixing strongly affects the isomer (interband) transitions being of crucial
importance for their existence, it has only a small impact on the g.s. MDM and a completely
negligible effect on the isomer MDM. The latter is explained by the circumstance, also
checked numerically in wave function (\ref{wtcoriol}), that while the g.s. is mixed with the
$I=5/2^{+}$ isomer-based state, the isomer state has no $I=3/2^{+}$ mixing-counterpart in
the g.s. band due to the $K\leq I$ restriction.
Thus, it appears that the connection between the ground and isomer state is mainly due to the
admixture in the g.s. This effect enters the transition matrix elements and   appears in the second power in the transition rates. We note that as the Coriolis mixing and the $g_R$
quenching counteract in the isomer transition rates and act in the same direction suppressing
$\mu_{\mbox{\scriptsize GS}}$ it may be possible to bring the latter closer to the
experimental value without drastic change in the predicted transition rates. However, this
would cause an overall change in all model observables, energy levels, $B(E2)$ and $B(M1)$
values, and would rather call for a full readjustment of the model parameters. Further
refinements such as more precise tuning of the QO deformation parameters entering DSM,
possible involvement of hexadecapole deformation (suggested in Ref.~\cite{Bemis_pscr88}),
or tuning of the spin-gyromagnetic quenching (similarly to $q_R$) as suggested in
Ref.~\cite{Ton70} could provide an even more reliable prediction of $^{229m}$Th
electromagnetic decay properties.

In conclusion, the CQOM-DSM-BCS model of the QPD spectrum and $B(M1)$ and
$B(E2)$ transition rates in $^{229}$Th provides a good description of the MDM
in the isomeric state. Our prediction $\mu_{\mbox{\scriptsize IS}}= -0.35
\mu_N$ is in good agreement with the experimentally determined value and
largely differs from the previous prediction in Ref.~\cite{Dyk98} based on
the Nilsson model. In the same time, the ground state MDM is overestimated by
a factor of approx. 1.4. We conclude that our results provide ground
for further refined consideration of the interplay between the nuclear
dynamic modes which determine the electromagnetic properties of the isomeric
state as a part of the entire $^{229}$Th structure.

The authors would like to thank Christoph H. Keitel for fruitful discussions.
This work is supported by the BNSF under contract DFNI-E02/6. AP
gratefully acknowledges funding by the EU FET-Open project 664732.


\bibliography{refs}

\end{document}